\documentclass[10pt, 3p]{elsarticle}

\usepackage{amsmath}

\usepackage[colorlinks]{hyperref}

\usepackage{enumitem}
\setlist[description]{labelindent=0pt, leftmargin=\parindent, style=nextline}

\usepackage[hyperref]{xcolor}
\definecolor{lifex}{HTML}{f60248}
\newcommand{\lifex}{\texorpdfstring{\texttt{life\textsuperscript{\color{lifex}{x}}}}{lifex}}

\definecolor{codegreen}{HTML}{009900}
\definecolor{codegray}{HTML}{808080}
\definecolor{codepurple}{HTML}{9400d1}
\definecolor{backcolor}{HTML}{f2f2eb}
\definecolor{linkcolor}{HTML}{0077b6}

\usepackage{tipa}
\usepackage{listings}

\lstdefinestyle{mystyle}{
    backgroundcolor=\color{white},
    commentstyle=\color{codegreen},
    keywordstyle=\color{lifex},
    numberstyle=\tiny\color{codegray},
    stringstyle=\color{codepurple},
    basicstyle=\ttfamily\footnotesize,
    breakatwhitespace=false,
    breaklines=true,
    captionpos=b,
    keepspaces=true,
    numbers=none,
    numbersep=5pt,
    showspaces=false,
    showstringspaces=false,
    showtabs=false,
    tabsize=4,
    frame=single
}

\lstdefinelanguage{prm}{
    keywords={subsection, set, end},
    comment=[l]{\#}
}

\lstdefinelanguage{xml}{
    keywords={value, default\_value, documentation, pattern, pattern\_description}
}

\lstdefinelanguage{json}{
    keywords={value, default\_value, documentation, pattern, pattern\_description}
}

\journal{SoftwareX}

\begin{document}
\hypersetup{allcolors=linkcolor}

\begin{frontmatter}
\title{\texorpdfstring{\lifex{}}{lifex}: a flexible, high performance library for the\texorpdfstring{\protect\\}{ }numerical solution of complex finite element problems}

\author{P. C. Africa}
\ead{pasqualeclaudio.africa@polimi.it}
\address{MOX, Department of Mathematics, Politecnico di Milano\protect\\
         Piazza Leonardo da Vinci, 32, 20133, Milano (Italy)}

\begin{abstract}
Numerical simulations are ubiquitous in mathematics and computational science. Several industrial and clinical applications entail modeling complex multiphysics systems that evolve over a variety of spatial and temporal scales. This study introduces the design and capabilities of \lifex{}, an open source \texttt{C++} library for high performance finite element simulations of multiphysics, multiscale, and multidomain problems. \lifex{} meets the emerging need for versatile, efficient computational tools that are easily accessed by users and developers. We showcase its flexibility and effectiveness on a number of illustrative examples and advanced applications of use and demonstrate its parallel performance up to thousands of cores.
\end{abstract}


\end{frontmatter}

\begin{table}[ht]
\centering
\begin{tabular}{|l|p{6.5cm}|p{6.5cm}|}
\hline
\textbf{Nr.} & \textbf{Code metadata description} & \textbf{Please fill in this column} \\
\hline
C1 & Current code version & v1.5.0 \\
\hline
C2 & Permanent link to code/repository used for this code version & Homepage: \url{https://lifex.gitlab.io/}, Code repository: \url{https://gitlab.com/lifex/lifex} \\
\hline
C3 & Code Ocean compute capsule & N/A \\
\hline
C4 & Legal Code License   & \href{https://www.gnu.org/licenses/lgpl-3.0.html}{LGPLv3} \\
\hline
C5 & Code versioning system used & \texttt{git} \\
\hline
C6 & Software code languages, tools, and services used & \texttt{C++} (standard \(\geq 17\)), \texttt{MPI}, \texttt{CMake} \(\geq 3.12.0\) \\\hline
C7 & Compilation requirements, operating environments, and dependencies & \texttt{deal.II} \(\geq 9.3.0\), \texttt{VTK} \(\geq 9.0.0\), \texttt{Boost} \(\geq 1.76.0\) \\
\hline
C8 & Link to developer documentation/manual & \url{https://lifex.gitlab.io/lifex/} \\
\hline
C9 & Support email for questions & \href{mailto:pasqualeclaudio.africa@polimi.it}{pasqualeclaudio.africa@polimi.it} \\
\hline
\end{tabular}
\caption{\lifex{} metadata.}
\label{tab:metadata}
\end{table}

\section{Motivation and significance}

A broad range of applications in biology, medicine, physics, engineering, astronomy, energy, environmental, and material sciences can be described by multiple physical processes interacting at different spatial and temporal scales \cite{Groen2014}. 
From the mathematical modeling perspective, such systems can be viewed as agglomerations of well-defined physics referred to as \textit{core models}. This explains the emerging need to develop new universal computational frameworks for the numerical simulation of multiphysics, multiscale, and multidomain problems. Such tools should easily enable the realization of \textit{in silico} experiments and provide a stable and intuitive development environment without compromising computational accuracy and efficiency.

The development of a tool of this kind plays a central role in decoupling the software development phase from the time-consuming process of performing different analyses -- from forward simulations to sensitivity analysis, optimization, and uncertainty quantification -- and in enabling the simulation of each core model in both standalone and various coupled configurations \cite{Keyes2013}.

\begin{figure}
    \centering
    \includegraphics[width=0.4\linewidth]{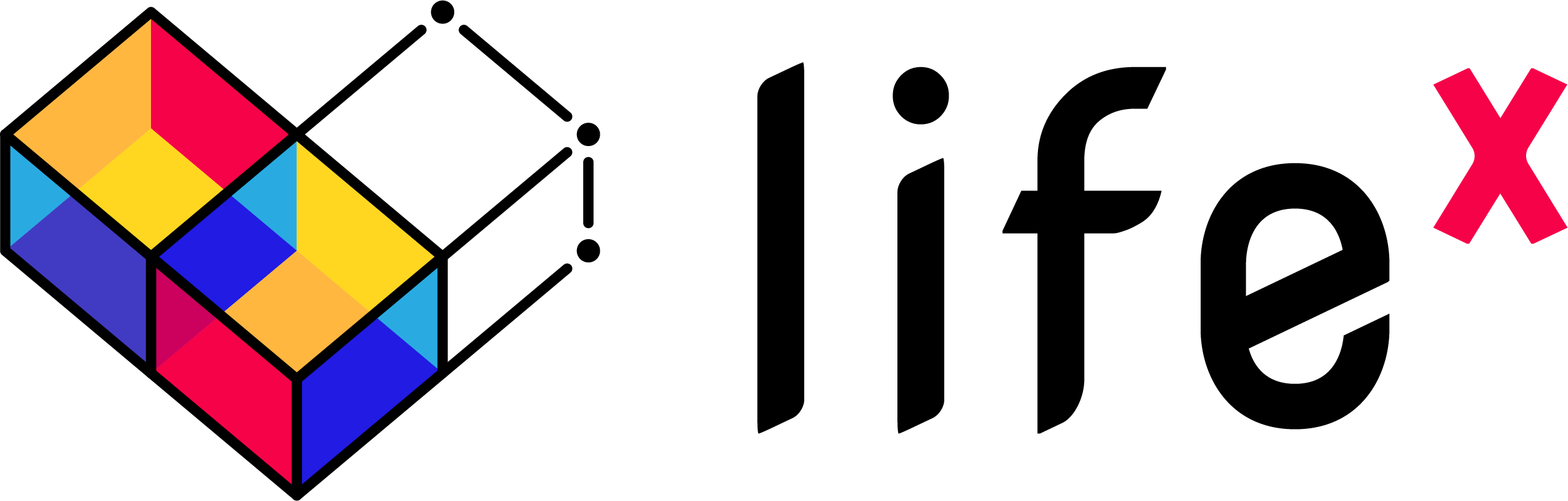}
    \caption{\lifex{} official logo. This image is licensed under a \href{https://creativecommons.org/licenses/by-sa/4.0/}{Creative Commons Attribution-ShareAlike 4.0 International License}.}
    \label{fig:logo}
\end{figure}

We introduce \lifex{} (pronounced \textipa{/\textsecstress la\textsci f\textprimstress\textepsilon ks/}, official logo shown in Fig.~\ref{fig:logo}), an open source library for the numerical solution of partial differential equations (PDEs) and related coupled problems, released under the \href{https://www.gnu.org/licenses/lgpl-3.0.html}{LGPLv3} license terms. It is written in \texttt{C++} using modern programming techniques available in the \texttt{C++17} standard and builds on the \texttt{deal.II} \cite{Arndt2021} finite element (FE) core. \lifex{} aims at providing a flexible and intuitive but robust and high performance tool simplifying the definition of complex physical models and their parameters, coupling schemes, and post-processing.

\lifex{} enables its users to shift the focus from technical numerics and implementation details toward the plain discrete mathematical formulation of the problems of interest. The library comes with extensive documentation and several examples and test cases that cover a wide range of applications and numerical strategies.

While serving similar purposes as existing multiphysics libraries in the open source community, such as \texttt{FEniCS} \cite{Alnaes2015,Scroggs2022}, \texttt{MFEM} \cite{Anderson2021}, \texttt{MOOSE} \cite{Permann2020}, and \texttt{preCICE} \cite{Bungartz2016}, \lifex{} offers several distinctive features, including the following:
\begin{itemize}
    \item an intuitive user programming interface with extreme ease of use;
    \item modern programming paradigms by design, leveraging the \texttt{C++17} standard, and up-to-date versions of third-party dependencies;
    \item parallel scalability up to thousands of cores;
    \item interoperability; that is, the possibility of importing and exporting data and meshes with common file formats, with particular reference to \href{https://vtk.org/}{\texttt{VTK}};
    \item support for arbitrary FEs, among those available in the \texttt{deal.II} backend \cite{Arndt2021a};
    \item the possibility to import meshes with either hexahedral or tetrahedral elements \cite{Quarteroni2008,Arndt2021};
    \item a clean and meticulously documented code base.
\end{itemize}
Each of these features is outlined below.

\section{Software description}
\label{sec:software_description}
\lifex{} was conceived in 2019 as an academic research library within the framework of the \href{https://iheart.polimi.it/}{iHEART} project (see \nameref{sec:acknowledgements}) at the Politecnico di Milano, with a primary focus on mathematical models and numerical schemes for integrated simulations of cardiac function.

Since its initial design, many modules for the simulation of different core models have been added to the code base. The development of \lifex{} was founded on strict coding conventions and practices \cite{Wilson2014}. The rapid increase in the number of developers and users testifies to the shallow learning curve of its kernel; it is fast and general enough to be used for diverse applications and merits being released as a standalone library.

Third-party dependencies of \lifex{} include the following:
\href{https://www.dealii.org/}{\texttt{deal.II}} (configured with support to \href{https://www.mcs.anl.gov/petsc/}{\texttt{PETSc}} \cite{Balay2022} and
\href{https://trilinos.github.io/}{\texttt{Trilinos}} \cite{Heroux2005}),
\href{https://vtk.org/}{\texttt{VTK}}, and
\href{https://www.boost.org/}{\texttt{Boost}}. \lifex{} can be configured to use, by default, linear algebra data structures and algorithms from \texttt{PETSc}, \texttt{Trilinos} (either through the interfaces exposed by \texttt{deal.II} or directly) or \texttt{deal.II} itself; where needed, a specific datatype or solver provided by one of the three backends may also be hard-coded, disregarding the default type with which \lifex{} was configured. All the code is natively parallel through the message passing interface (\texttt{MPI}); following a distributed memory paradigm, the global mesh is partitioned so that each \texttt{MPI} process owns and stores only a subset of cells.

This library aspires to maximum portability, having being deployed successfully on \texttt{Linux}, \texttt{Windows}, and \texttt{macOS} operating systems. This has motivated the use of advanced deployment technology, more specifically \href{https://github.com/elauksap/mk}{\texttt{mk}} \cite{Africa2022mk} (a set of portable, pre-compiled scientific packages for \texttt{x86-64} \texttt{Linux} systems), \href{https://gitlab.com/lifex/lifex-env}{\lifex{}\texttt{-env}} \cite{Africa2022lifex_env} (a set of build-from-source shell scripts explicitly inspired by \href{https://github.com/dealii/candi}{\texttt{candi}}), and \href{https://spack.io/}{\texttt{Spack}}. Pre-built \href{https://www.docker.com/}{\texttt{Docker}} images with all dependencies installed are also ready for download and use. More details can be found on the \href{https://lifex.gitlab.io/lifex/}{\lifex{} documentation}.

\subsection{Software architecture}
Structurally, the key features of \lifex{} can be grouped into three main components:
\begin{enumerate}
    \item An \textbf{abstraction layer} built on top of the \texttt{deal.II} FE library, exposing abstract numerical \textit{helpers} as essential building blocks that foster the development of advanced data structures and numerical schemes for time integration, linearization, solving and preconditioning linear systems, imposing boundary conditions, and mesh handling.
    \item A framework for \textbf{multiphysics coupling}, with functionalities enabling the transfer of solution fields and data from one core model to the other, either in the same domain or across multiple domains.
    \item A seamless \textbf{user interface} through several advanced input/output (I/O) capabilities, with a focus on importing data coming from the post-processing of experimental results, imaging techniques, or other numerical simulations, such as with the help of the \texttt{VTK} library.
\end{enumerate}

\begin{figure}
    \centering
    \includegraphics[width=0.75\linewidth]{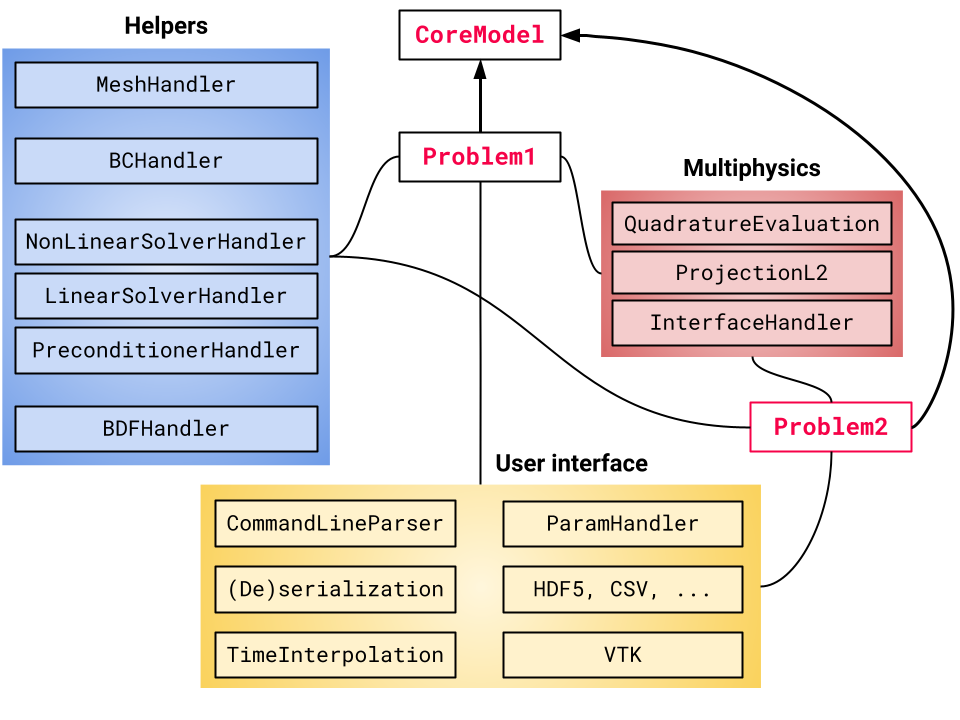}
    \caption{Overview of main \lifex{} components. The main classes and their interactions are shown, grouped into three categories: abstract numerical helpers (blue), multiphysics coupling (red), and user interface (yellow).}
    \label{fig:core}
\end{figure}

The main code components falling into these three categories, their classes, and their interactions are schematized in Fig.~\ref{fig:core}.

\subsection{Software functionalities}
All \lifex{} executables are built on the functionalities described below and are classified as follows:
\begin{description}
\item[\texttt{apps}:] Generic applications that are not model-specific, such as tools for printing mesh statistics or converting between compatible file formats.
\item[\texttt{examples}:] Problems and solvers that define specific model or geometric parameters, such as boundary conditions, initial conditions, domain, and so on.
\item[\texttt{tests}:] Executables used for automatic testing (run via \href{https://cmake.org/cmake/help/book/mastering-cmake/chapter/Testing%20With%20CMake%20and%20CTest.html}{\texttt{CTest}}),
automatically run on continuous integration (CI) services at each \texttt{git push} on \href{https://gitlab.com/}{\texttt{GitLab}} remote. Tests also include a number of \textit{tutorials}, which can be used as prototypes for building new applications. All tests and tutorials are used to determine the overall code coverage; that is, a metric that determines the number of lines of code that are successfully validated by the testing procedure.
\end{description}

A \lifex{} executable is typically associated with a set of common attributes, such as user-specified command line flags, the name of a parameter file (that is, a file containing all configurations, parameters, and settings used to run the executable, organized in a tree-like subsection structure), an \textit{execution mode} flag that specifies whether to generate a new parameter file or actually execute the app, an output directory containing all output files, \texttt{MPI} rank and size used for parallel computations. Upon running, such attributes are shared among instances of all classes. Moreover, all main classes are designed so as to expose their own specific parameters, such as geometry, physical parameters, discretization schemes, numerical settings, and I/O options, from the parameter file, each within its own subsection path.

The following three main classes define the minimal kernel interface common to all \lifex{} modules and executables:
\begin{description}
\item[\texttt{Core}:] A class implemented following the \textit{singleton} design pattern \cite{DesignPatterns1995} that stores attributes that are global and common to all other classes, such as those listed directly above.
\item[\texttt{CoreModel}:] An abstract class that inherits \texttt{Core} and extends it with pure virtual methods that define the interface exposed by each core model or numerical solver throughout \lifex{}. Classes in the \texttt{CoreModel} hierarchy expose a set of parameters that configure their behavior. Such parameters are exposed to the user through the parameter file (see Sec.~\ref{sec:user_interface}). A sample code snippet is provided and discussed in Sec.~\ref{sec:sample_code}.
\item[\texttt{lifex\_init}:] A lifespan handler that takes care of properly initializing all attributes and dependencies needed by each run, such as the instance of the singleton \texttt{Core} and \texttt{MPI}; an instance of this class is typically constructed at the very beginning of the \texttt{main()} function and destroyed at the program's end.
\end{description}

More specific high-level data structures are introduced below.

\subsubsection{Abstract numerical helpers}
An enormous part of \lifex{} consists of abstract wrappers and helpers: most of these classes explicitly invoke or refer to \texttt{deal.II} design and features \cite{Arndt2021a}, with the goal of exposing a higher-level interface to them and facilitating the implementation of advanced numerical schemes for a given problem. The main classes are described below.

\begin{description}
\item[\texttt{MeshHandler}:] A wrapper around \texttt{deal.II} distributed meshes.
The user can select whether to import a mesh with hexahedral or tetrahedral elements; depending on that choice, this class owns an instance of a \texttt{distributed} or a \texttt{fullydistributed} triangulation from \texttt{deal.II}; the latter is a recent introduction that adds support to tetrahedral meshes \cite{Arndt2021}, whose functionalities at the time of writing are still to be consolidated. The \texttt{MeshHandler} class interacts closely with \texttt{MeshInfo}, which parses information from the input mesh like volume and surface tags to be used, for example, to impose different boundary conditions on different parts of the boundary or to differentiate material properties in different sub-regions. Helper functions implemented in the \texttt{geometry/mesh\_info} and \texttt{geometry/finders} modules allow the computation of (sub)domain volumes and boundary surfaces or to locate, for example, the closest degree of freedom (DoF), mesh vertex, or boundary face to a given input point.

\item[\texttt{BCHandler}:] A helper class to impose different types of boundary conditions. Dirichlet boundary conditions can be either applied directly to an FE vector or imposed as linear constraints to the linear system arising from an FE discretization. For vector problems, normal or tangential fluxes can also be imposed. A helper method to assemble Neumann and Robin-like contributions to a local system's right-hand side is also provided.

\item[\texttt{LinearSolverHandler}:] For a sparse, distributed linear system, this class provides a simple interface that enables the user to select at run time which linear solver to use and all of its options (for instance, maximum number of iterations, tolerances, stopping criteria, and history log), as parsed from the parameter file. Many common solvers are included, such as \texttt{CG}, \texttt{GMRES}, \texttt{BiCGStab}, \texttt{MinRes}, \texttt{FGMRES}, but in principle any solver exposed by \texttt{deal.II} (including those from \texttt{PETSc} and \texttt{Trilinos}) is supported. Furthermore, the complete suite of solvers from \texttt{PETSc} remains accessible via the \texttt{-options\_file} command line flag, forwarded from \lifex{} to \texttt{PETSc}.

\item[\texttt{PreconditionerHandler}:] Analogously to \texttt{LinearSolverHandler}, this class exposes parameters that are used for the preconditioning of linear systems. It supports many preconditioner types, such as algebraic multi-grid (AMG), block Jacobi, additive Schwarz (SOR, SSOR, block SOR, block SSOR, ILU, ILUT), and can easily be extended to support more.

\item[\texttt{BDFHandler}:] For time-dependent problems, semi-implicit backward difference formula (BDF) time discretization schemes \cite{Forti2015} are implemented in this class, which deals with storing the information to advance the problem from one time step to the next. This class stores and exposes the BDF solution and its extrapolation and can easily be extended to different time-advancing schemes.

\item[\texttt{NonLinearSolverHandler}:] For solving non-linear problems, a family of Newton methods is provided. An abstract implementation requires the user to specify an \textit{assemble function}, which assembles the Jacobian matrix and the residual vector, and a \textit{solve function} that assembles the preconditioner and solves the linear system associated with each non-linear iteration; the two functions must return the norms of residual, solution, and Newton increment to be used as possible stopping criteria. The \textit{frozen Jacobian} (or \textit{Jacobian lagging}) approach \cite{Brown2013},
which consists of reassembling the Jacobian only once every \(n\) time steps, can be toggled to increase computational efficiency. Two specializations for the quasi-Newton method with the Jacobian matrix approximated via finite differences \cite{Gill1972} and for the inexact Newton method \cite{Eisenstat1996} are also supplied. Moreover, each non-linear solution scheme can be equipped with proper acceleration strategies (static relaxation, Aitken extrapolation \cite{Kuettler2008}, and Anderson acceleration \cite{Walker2011}) to accelerate convergence.
In addition to the non-linear solver handler, the user can benefit from the use of \textit{automatic differentiation} (with support for the \href{https://trilinos.github.io/sacado.html}{\texttt{Sacado}} and \href{https://github.com/coin-or/ADOL-C}{\texttt{ADOL-C}} interfaces exposed by \texttt{deal.II}), demonstrated on \texttt{Tutorial04\_AD} and \texttt{Tutorial07\_AD}, which enables implementing the computation of exact derivatives (up to machine precision) of complicated functions very easily.
\end{description}

\subsubsection{Multiphysics coupling}
The complexity of multiphysics, multiscale, and multidomain problem can be relieved with the help of three hierarchies of classes that all serve the purpose of transferring solution fields and data either from one core model to another or across internal interfaces.

In order to keep the code as general as possible, we assume that different core models can be solved using arbitrarily independent discretization schemes, such as different FE degrees or mesh resolutions. This improves the capturing of all physical phenomena involved, even though their dynamics can be characterized by vastly different spatial and temporal scales.

\begin{figure}
    \centering
    \includegraphics[width=0.9\linewidth]{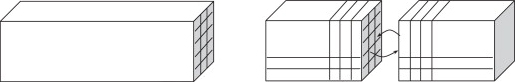}
    \caption{Possible solution schemes for a geometrically coupled problem: monolithic (left) vs. partitioned (right) solution scheme. Reprinted from \cite{Borgdorff2014}. The original image is licensed under a \href{https://creativecommons.org/licenses/by/3.0/}{CC BY 3.0 License}.}
    \label{fig:multidomain}
\end{figure}

We note that problems involving more than one physical model (possibly on multiple domains sharing a common interface, such as in the case of fluid-structure interaction) can be generally solved using either monolithic or partitioned algorithms \cite{Bucelli2021}, as schematized in Fig.~\ref{fig:multidomain}. In the former case, a global system involving all unknowns from all problems is assembled and solved at each time step; in the latter, each sub-problem is solved independently, and coupling conditions are imposed, for example by using explicit schemes or sub-iterating with a fixed-point scheme until a convergence of coupling conditions is reached. Both choices are possible in \lifex{} and illustrated by a number of examples and tests.

\begin{figure}
    \centering
    \includegraphics[width=0.75\linewidth]{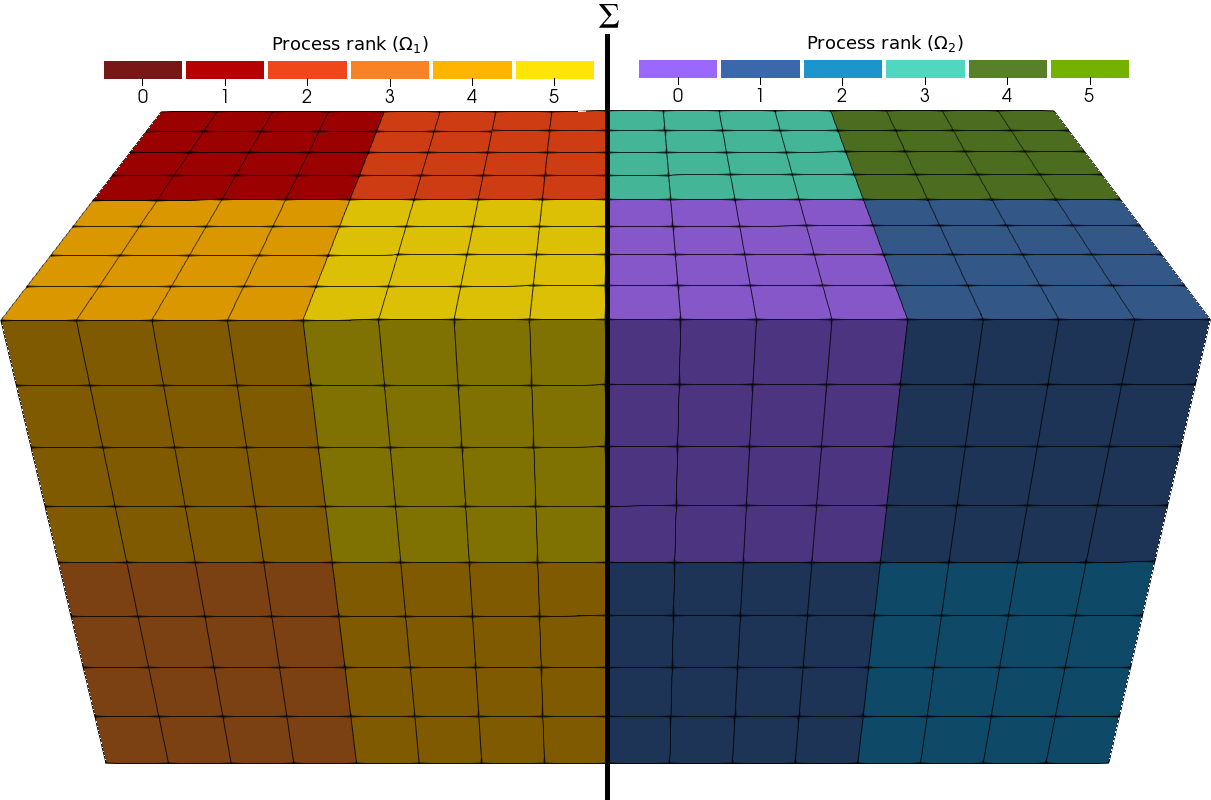}
    \caption{Example of handling two domains \(\Omega_1\) (left) and \(\Omega_2\) (right) sharing a common interface \(\Sigma\) with conforming mesh discretizations. The \texttt{InterfaceHandler} is able to deal properly with non-conforming parallel partitioning.}
    \label{fig:multidomain_partitioning}
\end{figure}

\begin{description}
\item[\texttt{QuadratureEvaluation}:] This class provides a high-level interface for the evaluation of arbitrary analytic functions or more complex data structures at a given quadrature point. User-defined classes deriving from \texttt{QuadratureEvaluation} can easily be implemented for scalar, vector, or tensor fields. Furthermore, the \texttt{QuadratureEvaluationFEM} hierarchy of classes is implemented to enable the coupling of multiple FE models solved in the same domain. Different problems can be discretized using different FE degrees, and the integrals arising from the weak formulation can be approximated using quadrature formulas of different types and degrees of accuracy. The \texttt{QuadratureEvaluationFEM} classes provide an interface similar to that of \texttt{FEValues} from \texttt{deal.II}: such objects are constructed using the \texttt{DoFHandler} associated with the FE field to be evaluated and the quadrature rule used for the target problem. By re-initializing such objects on each mesh cell, the input field can be evaluated at the corresponding quadrature points. \lifex{} provides specializations to automatically evaluate the FE solutions, gradients, and divergence of a given solution vector.

\item[\texttt{ProjectionL2}:] Instead of the exact numerical evaluation allowed by \texttt{QuadratureEvaluation} classes, a smoothed \(L^2\) projection can be considered. Given a function \(f(\textbf{x})\), this class computes a FE solution \(f_h(\textbf{x})\) that satisfies \((\varepsilon\nabla f_h, \nabla\varphi_i)_{\Omega}+(f_h, \varphi_i)_{\Omega} = (f, \varphi_i)_{\Omega}\) for each basis function \(\varphi_i\) in the chosen FE space. The numerical solution to this problem clearly involves a mass matrix: its lumping can be toggled, and the regularization parameter \(\varepsilon\) can be tuned to prevent numerical oscillations, for example in the case of coarse meshes \cite{Quarteroni2010}. The solution \(f_h\) obtained can thus easily be evaluated at the quadrature nodes associated with the target problem.

\item[\texttt{InterfaceHandler}:] Consider two subdomains \(\Omega_1\) and \(\Omega_2\) sharing a common interface \(\Sigma\) with conforming discretizations, and let \(u_1\) and \(u_2\) be FE functions defined on the two subdomains, typically representing solutions to differential problems defined on the two subdomains. Suppose that the problem defined on \(\Omega_1\) (\(\Omega_2\)) involves conditions on \(\Sigma\) that depend on \(u_2\) (\(u_1\)) \cite{Quarteroni1999,Bucelli2021}. \texttt{InterfaceHandler} builds the \textit{interface maps}; that is, two mappings of DoFs between the local interface \(\Sigma\) and the global domains \(\Omega_1\) and \(\Omega_2\). This is of critical importance in parallel simulations, where the parallel partitioning on both domains can be different, as in the example in Fig.~\ref{fig:multidomain_partitioning}. Finally, for each subdomain, this class manages the extraction of interface data on \(\Sigma\) from the other subdomain and its application as a boundary condition on \(\Sigma\). This class deals only with conforming meshes; extensions to non-conforming discretizations, such as  the \texttt{INTERNODES} technique \cite{Deparis2016}, are still under development.
\end{description}

The last case to be considered is transferring solutions between multiple core models solved using the same FE discretization but with different mesh resolutions. For nested hexahedral grids, the \texttt{VectorTools} namespace of \texttt{deal.II} already provides functions that perform precisely the interpolation needed. This procedure is hardly generalizable as it depends heavily on how the different meshes have been generated and on the mesh element type. For instance, transfer operators built on radial basis function (RBF) interpolators could be used in the case of non-conforming discretizations \cite{Deparis2014,Salvador2020} but have yet to be implemented.

Clearly, the two approaches can be combined to couple different models solved with both different FE approximations and mesh resolutions.

\subsubsection{User interface}
\label{sec:user_interface}

\begin{description}
\item[\texttt{CommandLineParser}:]
\lifex{} makes use of the lightweight parser \href{https://github.com/muellan/clipp}{\texttt{clipp}} for parsing command line arguments. All executables expose a set of command line options that can be printed using the \verb|-h| (or \verb|--help|) flag:
\begin{lstlisting}[language=bash]
./executable_name -h
\end{lstlisting}

\item[\texttt{ParamHandler}:]
Each \lifex{} executable defines a set of parameters that are required in order to be run. They involve problem-specific
parameters (such as coefficients, geometry, time interval,
and boundary conditions), numerical parameters (such as types of linear/non-linear solvers, tolerances, and maximum number of iterations), I/O options, and so on. In the event an application has sub-dependencies such as a linear solver, the related parameters are also included, typically in a proper subsection path. Parameters are organized in a tree-like structure following the functionalities exposed by the \texttt{ParameterHandler} class from \texttt{deal.II}. The first step before running any executable is to generate the default parameter file(s) via the \verb|-g| (or \verb|--generate-params|) flag:
\begin{lstlisting}[language=bash]
./executable_name -g -f filename.ext
\end{lstlisting}

At the user's option, in order to guarantee a flexible interface with external file processing tools, the parameter file extension \verb|ext| can be chosen among three different interchangeable file formats \verb|prm|, \verb|json| or \verb|xml|, sorted from the most human-readable to the most machine-readable.

An excerpt of a \verb|prm| file follows:
\begin{lstlisting}[language=prm, label=lst:prm, caption={Example of parameter file in \texttt{prm} format. The tree-like subsection structure is emphasized.}]
subsection Problem
  subsection Mesh and space discretization
    # Parameter description goes here.
    set Element type = Hex
    # ...
  end
  # ...
  subsection Linear solver
    set Type = GMRES
    
    subsection GMRES
      set Max. number of temporary vectors = 100
      # ...
    end
  end
  # ...
  subsection Preconditioner
    set Type = AMG
    
    subsection AMG
      set W-cycle = true
      # ...
    end
end
\end{lstlisting}

A parameter file can easily be set up using any text editor, without need to recompile the source code. Finally, omitting the \verb|-g| flag in the command above, an existing parameter file is read and the simulation subsequently run.

The \texttt{ParamHandler} class of \lifex{} extends the \texttt{deal.II} class by two main functionalities:
\begin{description}
    \item[verbosity control:] By default, only parameters declared to have a \textit{standard} verbosity are printed. In order to customize the user experience, the verbosity of each parameter can be decreased (\textit{minimal}) or increased (\textit{full}) from the source code. A parameter file containing a minimal (full) set of parameters can be generated by passing the optional flag \verb|minimal| (\verb|full|) to the \verb|-g| flag:
\begin{lstlisting}[language=bash]
./executable_name -g [minimal,full] \
                  -f filename.ext
\end{lstlisting}
    If the \verb|-g| is provided without any further specification, the intermediate level of verbosity is assumed.

    \item[multiple default values:] In principle, each application could be run to simulate different scenarios or simply with different predefined sets of parameters; \lifex{} offers the possibility of providing multiple default parameter files out of the box. The \texttt{ParamHandler} class can read user-provided files in \verb|json| format by specifying a list of parameter names and their (new) default values, which will be appended to the complete set of parameters and written to a ready-to-use file (see, for example, the \texttt{time\_interpolation} test).
\end{description}

Utilities for parsing lists of values are also provided in the \texttt{param\_handler\_helpers} module for convenience of use.

\item[(De-)serialization:]
\lifex{} includes a checkpointing system that allows for all aspects of a simulation to be serialized to file. This allows recovering a simulation state after an unexpected failure, restarting after maximum computational wall time has been reached, or simply initializing a simulation with custom input data. Convenient tools for (de-)serializing (distributed) meshes and solution vectors are provided in the \texttt{io/serialization} module, with an interface to \texttt{deal.II}-compatible binary files, and their use is demonstrated in the \texttt{serialization} test.

\item[\texttt{CSV} readers and writers:]
The simplicity of use of comma-separated value (CSV) files makes it a widely chosen option to process data organized into fields. Many utility functions and classes are present in \lifex{} to read and write \texttt{CSV} files by converting number and text values into \href{https://en.cppreference.com/w/cpp/container}{\texttt{STL} containers} or \texttt{deal.II} data structures (vectors, matrices, and so on). This enables easily post-processing simulation results, for example by exporting point-wise variables at each time step.

\item[\texttt{TimeInterpolation}:]
Many applications require resampling discrete sets of data at arbitrary points, such as time-dependent variables that need to be interpolated in correspondence with the time steps performed by the numerical simulation. The \texttt{TimeInterpolation} class provides methods based on linear interpolation, cubic splines, smoothing cubic splines, trigonometric interpolation (discrete Fourier transform), and linear and spline interpolation of the derivative of the input data.

\item[\texttt{VTKFunction} and \texttt{VTKPreprocess}:]
Many physical problems are characterized by coefficients derived from experimental data or imaging techniques, such as segmented geometries of organs from magnetic resonance imaging (MRI) or computer tomography (CT) scans \cite{Paliwal2021,Fumagalli2020}, or from post-processing of other numerical simulation steps \cite{Regazzoni2022}. The \href{https://vtk.org/}{\texttt{VTK}} toolkit defines some of the most common data formats to deal with data defined over volumes (\texttt{vtkUnstructuredGrids}) or surfaces (\texttt{vtkPolyData}). Moreover, it is also used in sophisticated pipelines for surface processing and mesh generation \cite{Fedele2021}. \lifex{} provides a class named \texttt{VTKFunction}, inherited from \texttt{dealii::Function}, that imports a \texttt{VTK} file containing a cell or point data field and evaluates it at an arbitrary point, possibly associated with a computational mesh. Three possible evaluation methods are available: closest point, linear projection, and signed distance. Finally, the \texttt{VTKPreprocess} class exploits \texttt{VTKFunction} to interpolate input \texttt{VTK} data onto FE vectors, which are serialized to file for later importing and reuse in numerical simulations.
\end{description}

\subsection{Sample code snippet}
\label{sec:sample_code}

The following code illustrates a sample code snippet with comments, containing the minimal interface exposed by the vast majority of all \lifex{} classes; that is, those inherited from \texttt{CoreModel}. In particular, the \verb|declare_parameters| and \verb|parse_parameters| methods are \textit{pure virtual} and must be overridden, whereas the \verb|run| method is virtual and has an empty definition by default. An example of how to locally adjust the verbosity of some parameters is also shown. Finally, this sample class makes use of a \texttt{LinearSolverHandler}, for which we also declare and parse related parameters.

\begin{lstlisting}[language=c++]
namespace lifex
{
  class Problem : public CoreModel
  {
  public:
    // Specify the subsection path where to
    // declare current parameters.
    Problem(const std::string &subsection_path)
    : CoreModel(subsection_path)

    // Specify a "relative" subsection.
    // Subpaths are separated by a "/".
    , linear_solver(
        prm_subsection_path + " / Linear solver",
        /* ... */)
    {}

    virtual void
    declare_parameters(ParamHandler &params) const override
    {
      // Navigate subsections and declare parameters.
      params.enter_subsection_path(prm_subsection_path);
      {
        // Problem-dependent parameters.
        // ...

        params.set_verbosity(VerbosityParam::Full);
        {
          // If -g full is *not* specified,
          // the parameters declared here will
          // be hidden from the parameter file.
          // ...
        }
        params.reset_verbosity();
      }
      params.leave_subsection_path();

      linear_solver.declare_parameters(params);
    }

    virtual void
    parse_parameters(ParamHandler &params) override
    {
      // Actually parse parameter file.
      params.parse();

      // Analogously to declare_parameters,
      // navigate subsections, read parameters,
      // and possibly store them into class members.
      // ...

      linear_solver.parse_parameters(params);
    }

    virtual void
    run() override
    {
      // Create mesh.
      // Setup system.
      // Assemble system.
      // Solve system.
      // Output solution.
    }

  private:
    LinearSolverHandler linear_solver;
    // ...
  };
}
\end{lstlisting}

\section{Illustrative examples}
\lifex{} is capable of solving complex multiphysics problems. The functionalities described in the previous section are pointed out in a series of \textit{tutorials} that are found in the source code as tests. The tutorials are sorted by increasing complexity and involve different kinds of scalar or vector equations and coupled problems, solved either monolithically or partitioned. Here, we provide a summary of the tutorials available and the corresponding PDEs solved.

\begin{description}
    \item[\texttt{Tutorial01}:] Linear elliptic equation:
    \begin{equation*}
    -\Delta u = f, \quad \text{in } (-1, 1)^3.
    \end{equation*}
    \item[\texttt{Tutorial02}:] Linear parabolic equation:
    \begin{equation*}
        \frac{\partial u}{\partial t} - \Delta u + u = f, \quad \text{in } (-1, 1)^3 \times (0, T].
    \end{equation*}
    \item[\texttt{Tutorial03}:] Non-linear elliptic equation:
    \begin{equation*}
        -\Delta u + u^2 = f, \quad \text{in } (-1, 1)^3.
    \end{equation*}
    \item[\texttt{Tutorial04}:] Non-linear parabolic equation:
    \begin{equation*}
        \frac{\partial u}{\partial t} - \Delta u + u^2 = f, \quad \text{in } (-1, 1)^3 \times (0, T].
    \end{equation*}
    \item[\texttt{Tutorial04\_AD}:] The same as \texttt{Tutorial04}, with the Jacobian matrix assembled via automatic differentiation.
    \item[\texttt{Tutorial05}] Parabolic system of equations, solved monolithically:
    \begin{equation*}
        \left\{
        \begin{aligned}
        \frac{\partial u}{\partial t} - \Delta u + u^2 &= f, & \quad & \text{in } (-1, 1)^3 \times (0, T], \\
        \frac{\partial v}{\partial t} - \Delta v + uv &= g, & \quad & \text{in } (-1, 1)^3 \times (0, T].
        \end{aligned}
        \right.
    \end{equation*}
    \item[\texttt{Tutorial06}:] The same as \texttt{Tutorial05}, solved using an explicit partitioned scheme and exploiting the \texttt{QuadratureEvaluationFEM} capabilities.
    \item[\texttt{Tutorial07}:] Cahn-Hilliard equation:
    \begin{equation*}
        \left\{ \begin{aligned} \frac{\partial c}{\partial t} - \Delta \mu &= 0, & \quad & \text{in } (0, 1)^3 \times (0, T], \\
        \mu - \frac{\mathrm{d}f}{\mathrm{d}c}(c) + \lambda \Delta c &= 0, & \quad & \text{in } (0, 1)^3 \times (0, T].
        \end{aligned}
        \right.
    \end{equation*}
\end{description}

For further details about the mathematical and numerical formulations of all these problems, such as boundary and initial conditions, please refer to the \href{https://lifex.gitlab.io/lifex/}{\lifex{} documentation}.

We present below three examples that showcase the main features of \lifex{}. All the results shown are new, original contributions.
First, we prove that the abstract helpers for the advanced numerical schemes described in Sec.~\ref{sec:software_description} do not affect parallel performance, as the speedup is almost approximately linear up to thousands of cores; then, we present a multidomain problem where two Stokes problems are solved on two cubes sharing a common face with proper interface conditions, proving that monolithic and partitioned schemes for domain decomposition problems can easily be implemented with a negligible computational overhead due to the parallel transfer of solutions across the interface; finally, an advanced, fully implicit numerical solver for the Cahn-Hilliard equation demonstrates the ease of implementation and the enormous flexibility available to users when dealing with complex multiphysics problems.

\subsection{Scalability study}
\label{sec:speedup}
We perform a strong scaling test on \texttt{Tutorial06}, where the following equations are solved:
\begin{equation*}
\left\{
\begin{aligned} \frac{\partial u}{\partial t} - \Delta u + u^2 &= f, & \quad & \text{in } \Omega \times (0, T] = (-1, 1)^3 \times (0, T], \\
\frac{\partial v}{\partial t} - \Delta v + uv &= g, & \quad & \text{in } \Omega \times (0, T], \\
u &= u_\mathrm{ex}, & \quad & \text{on } \partial\Omega \times (0, T], \\
v &= v_\mathrm{ex}, & \quad & \text{on } \partial\Omega \times (0, T], \\
u &= u^0, & \quad & \text{in } \Omega \times \{0\}, \\
v &= v^0, & \quad & \text{in } \Omega \times \{0\},
\end{aligned}
\right.
\end{equation*}
where \(f, g, u^0\), and \(v^0\) are chosen such that the exact solution is
\begin{equation*}
    \left\{
    \begin{aligned} u_\mathrm{ex}(\mathbf{x}, t) &= t \cos(\pi x_0) \cos(\pi x_1) \cos(\pi x_2), \\
    v_\mathrm{ex}(\mathbf{x}, t) &= e^t \lVert\mathbf{x}\rVert^2.
    \end{aligned}
    \right.
\end{equation*}

The two equations are discretized in time using the \texttt{BDFHandler} of order \(1\) for \(u\) and \(3\) for \(v\), decoupled using an explicit partitioned scheme, and linearized using the \texttt{NonLinearSolverHandler} class. Finally, the FE space discretization consists of linear (quadratic) elements for \(u\) (\(v\)). The solution \(u\) appearing in the second equation is evaluated using the capabilities of \texttt{QuadratureEvaluationFEM}. The mesh size consists of 2,097,152 cells (average cell diameter: \(h \approx 0.027\)) and 19,121,282 DoFs (2,146,689 for \(u\), 16,974,593 for \(v\)), the time step chosen is equal to \(\Delta t = 0.1\), and the simulation is run until \(T=1\).

\begin{figure}
    \centering
    \includegraphics[width=0.5\linewidth]{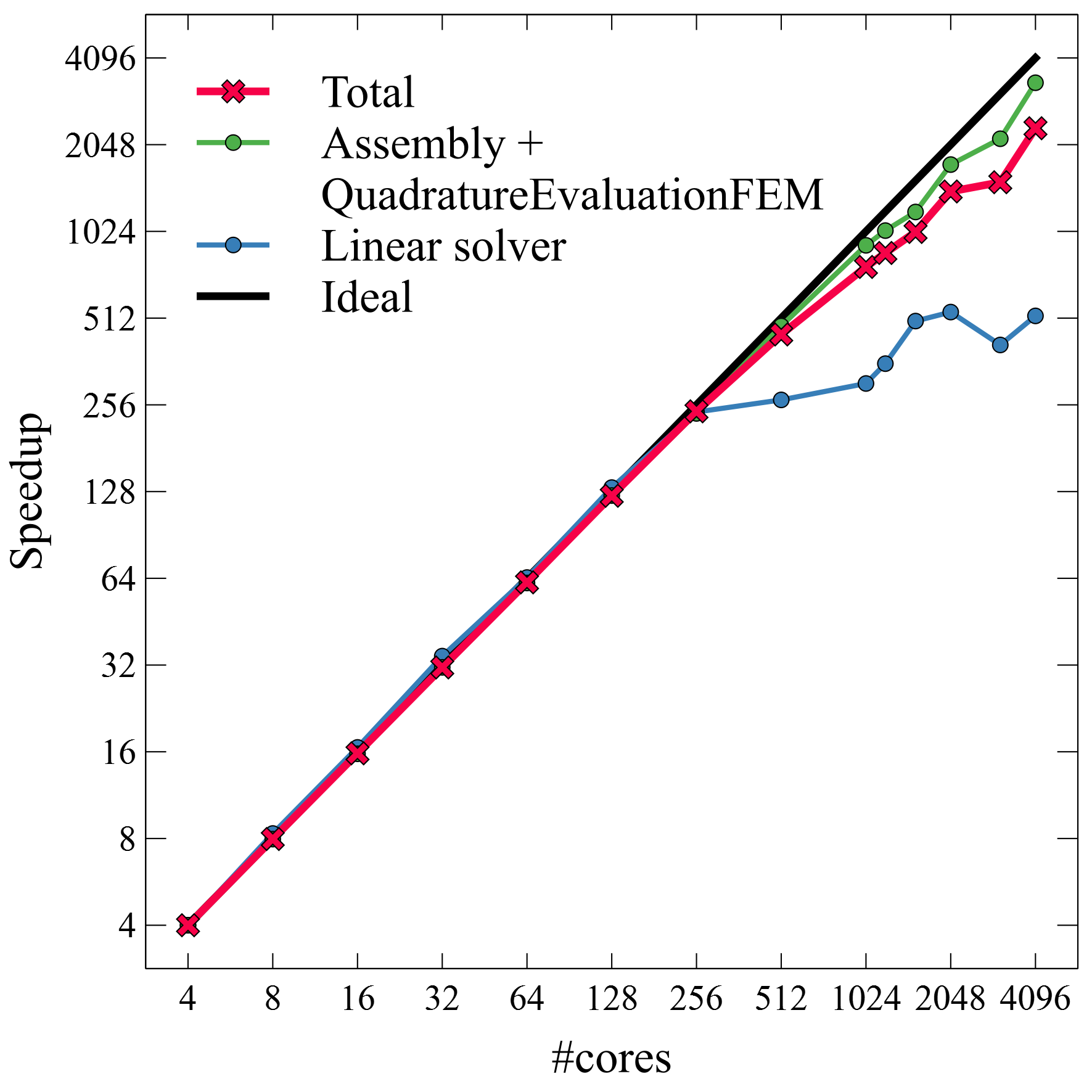}
    \caption{Parallel speedup of \lifex{}, demonstrated on \texttt{Tutorial06}. The speedup was computed on total time (red), time spent in assembling the linear system (including the evaluation of the \texttt{QuadratureEvaluationFEM} field) at each time step (green), and time spent in solving the linear system at each time step through the \texttt{LinearSolverHandler}, together with the \texttt{PreconditionerHandler} wrappers (blue).}
    \label{fig:speedup}
\end{figure}

The scalability test was run on the \href{https://wiki.u-gov.it/confluence/display/SCAIUS/UG3.3%3A+GALILEO100+UserGuide}{\texttt{GALILEO100}}
supercomputer available at \texttt{CINECA} (Intel CascadeLake 8260, 2.40GHz). We recorded the total simulation time and partial times spent in the assembly and linear-solving phases; the speedup for the three quantities shown in Fig.~\ref{fig:speedup} confirms that on such a benchmark problem, the main \lifex{} data structures scale approximately linearly up to \(4,096\) cores. The linear solver performances slowly degrades beginning at about \(512\) cores, which is likely due to the limited problem size. Table~\ref{tab:quadrature_evaluation} reports the summary of the computational costs for the different phases of a run of \texttt{Tutorial06} on \(1024\) cores.

Some interesting conclusion can be drawn. First, the evaluation of \(u\) at quadrature nodes of the FE space used for the discretization of the equation for \(v\) is invoked \(\approx 443\) million times, which makes a substantial contribution to the assembly phase (about \(27\%\) of the total time): nevertheless, as Fig.~\ref{fig:speedup} shows, the assembly phase still scales almost perfectly linearly, which proves that the implementation of the \texttt{QuadratureEvaluation} hierarchy of classes introduces a computational overhead that scales almost ideally in parallel. Moreover, the additional overhead from applying the \texttt{BDFHandler}, \texttt{NonLinearSolverHandler}, \texttt{LinearSolverHandler}, and \texttt{PreconditionerHandler} wrappers is negligible and does not affect the solver's overall parallel performance. This shows that \lifex{} can reach an ideal parallel speedup while the abstract numerical helpers and multiphysics coupling interface enable a significant reduction in the total number of lines of code compared to a naive implementation based only on \texttt{deal.II}, thus letting the user focus on the plain discrete formulation of the problems of interest rather than on technical numerics and implementation details.

\begin{table}
    \centering
    \begin{tabular}{l|r|r|r}
        \textbf{Section} & \textbf{No. calls} & \textbf{Wall time} & \textbf{\% of total} \\\hline
        Solver for \(u\): solve time step & 11 & 31.515s & 3.05\% \\
        Solver for \(u\): non-linear solver & 11 & 29.460s & 2.85\% \\
        Solver for \(u\): preconditioner assembly + linear solver & 33 & 25.907s & 2.51\% \\
        Solver for \(u\): system assembly & 44 & 3.365s & 0.33\% \\
        Solver for \(u\): linear solver & 33 & 1.799s & 0.17\% \\\hline
        Solver for \(v\): solve time step & 11 & 988.561s & 95.79\% \\
        Solver for \(v\): system assembly & 11 & 956.339s & 92.67\% \\
        Solver for \(v\): \texttt{QuadratureEvaluationFEM} initialization & 11 & 0.000s & 0.000\% \\
        Solver for \(v\): \texttt{QuadratureEvaluationFEM} re-initialization & 22,528 & 0.020s & 0.000\% \\
        Solver for \(v\): \texttt{QuadratureEvaluationFEM} evaluation & 443,418,624 & 278.760s & 27.32\% \\
        Solver for \(v\): preconditioner assembly + linear solver & 11 & 23.601s & 2.29\% \\
        Solver for \(v\): linear solver & 11 & 13.897s & 1.35\% \\\hline
        Total wallclock time & & 1031.991s & 100\%
    \end{tabular}
    \caption{Summary of computational costs of a run of \texttt{Tutorial06} on \(1024\) cores.}
    \label{tab:quadrature_evaluation}
\end{table}

\subsection{Multidomain problems}
The \texttt{multidomain\_stokes} example was run to demonstrate the parallel performance of the multidomain capabilities of \lifex{}, with particular reference to the \texttt{InterfaceHandler} class.

The problem being solved is the Stokes model:
\begin{equation*}
    \left\{
    \begin{alignedat}{2} -\mu \Delta \mathbf{u} + \nabla p & = 0, & \quad & \text{in }\Omega, \\
    \nabla\cdot\mathbf{u} & = 0, & \quad & \text{in }\Omega, \\
    \mathbf{u} & = [1, 0, 0]^T, & \quad & \text{on }\Gamma_{\mathrm{in}}, \\
    \mathbf{u} & = [0, 0, 0]^T, & \quad & \text{on }\Gamma_{\mathrm{sides}}, \\
    \mu \nabla \mathbf{u} \cdot \mathbf{\nu} - p\mathbf{n} & = 0, & \quad & \text{on }\Gamma_{\mathrm{out}},
    \end{alignedat}
    \right.
\end{equation*}
where
\begin{equation*}
    \begin{aligned}
    \Omega &= (-0.5, 1.5)\times(-0.5, 0.5)\times(-0.5, 0.5),\\
    \Gamma_{\mathrm{in}} &= \left\{ x = -0.5 \right\},\\
    \Gamma_{\mathrm{out}} &= \left\{ x = 1.5 \right\},\\
    \Gamma_{\mathrm{sides}} &= (\partial\Omega\backslash(\Gamma_{\mathrm{in}}\cup\Gamma_{\mathrm{out}}))^{\mathrm{o}},
    \end{aligned}
\end{equation*}
and \(\mathbf{\nu}\) denotes the outward unit normal. \(\Omega\) is split into the two subdomains \(\Omega_0\) and \(\Omega_1\) across the interface \(\Sigma\), which is defined as
\begin{equation*}
    \begin{aligned}
    \Omega_0 &= (-0.5, 0.5)\times(-0.5, 0.5)\times(-0.5, 0.5),\\
    \Omega_1 &= (0.5, 1.5)\times(-0.5, 0.5)\times(-0.5, 0.5),\\
    \Sigma &= \left\{x = 0.5\right\}.
    \end{aligned}
\end{equation*}

Denoting the solution on the subdomain \(\Omega_i\) for \(i=0,1\) by \(\mathbf{u}_i,\, p_i\), the multidomain formulation of the problem is as follows:
\begin{equation*}
    \left\{
    \begin{alignedat}{2}
    -\mu \Delta \mathbf{u}_i + \nabla p_i & = 0, & \quad & \text{in }\Omega_i,\\
    \nabla\cdot\mathbf{u}_i & = 0, & \quad & \text{in }\Omega_i, \\
    \mathbf{u}_0 & = [1, 0, 0]^T, & \quad & \text{on }\Gamma_{\mathrm{in}}, \\
    \mathbf{u}_i & = [0, 0, 0]^T, & \quad & \text{on }\Gamma_{\mathrm{sides}}\cap\partial\Omega_i, \\
    \nabla \mu \mathbf{u}_1 \mathbf{\nu} - p\mathbf{\nu} & = 0, & \quad & \text{on }\Gamma_{\mathrm{out}}, \\
    \mathbf{u}_0 & = \mathbf{u}_1, & \quad & \text{on }\Sigma, \\
    -\mu \nabla \mathbf{u}_0 \mathbf{\nu}_0 - p_0 \mathbf{\nu}_0, & = \mu \nabla \mathbf{u}_1 \mathbf{\nu}_1 + p_1 \mathbf{\nu}_1 & \quad & \text{on }\Sigma,
    \end{alignedat}
    \right.
\end{equation*}
where the two conditions on \(\Sigma\) denote the continuity of velocity and stresses across the interface.

This problem can be solved by using either a fixed-point or a monolithic scheme. The goal of this section is to demonstrate the performance of the \texttt{InterfaceHandler} class. We thus ran a simulation using 6,714,692 DoFs on each subdomain \(\Omega_i\) for \(i=0,1\) (6,440,067 for the velocity and 274,625 for the pressure block), resulting in a total number of DoFs at the interface \(\Sigma\) equal to 49,923. The simulation requires 40 fixed-point iterations to meet the prescribed tolerance of \(10^{-6}\) on the increment norm of the interface data \(\mathbf{u}_1\).

Table~\ref{tab:multidomain_stokes} reports the summary of the computational costs for the different phases of a run of the \texttt{multidomain\_stokes} example on \(1024\) cores. As the number of interface DoFs is much smaller than the total number of volume DoFs, the time spent in setting up interface maps and transferring the solutions between the two subdomains \(\Omega_0\) and \(\Omega_1\) is negligible with respect to the phases of assembling and solving the associated linear systems. This demonstrates that the overall computational cost associated with a multidomain simulation, such as in a fluid-structure interaction (FSI) framework, is largely dominated by the time spent in assembling and solving the associated linear system(s), whereas the overhead introduced by the \texttt{InterfaceHandler} class is negligible.

As a consequence, the speedup results shown in Sec.~\ref{sec:speedup} still hold for multidomain problem, solved either monolithically or partitioned, as long as the parallel partitioning of \(\Omega_0\) and \(\Omega_1\) is fairly load-balanced.

\begin{table}
    \centering
    \begin{tabular}{l|r|r|r}
        \textbf{Section} & \textbf{No. calls} & \textbf{Wall time} & \textbf{\% of total} \\\hline
        \texttt{InterfaceHandler}: build interface maps & 1 & 24.1s & 0.17\% \\
        \texttt{InterfaceHandler}: extract interface values & 80 & 2.38s & 0\% \\
        \texttt{InterfaceHandler}: apply Dirichlet interface conditions & 80 & 0.875s & 0\% \\\hline
        Problem in \(\Omega_0\): preconditioner assembly + linear solver & 40 & 6.73e+03s & 48\% \\
        Problem in \(\Omega_0\): system assembly & 40 & 42.2s & 0.3\% \\
        \hline
        Problem in \(\Omega_1\): preconditioner assembly + linear solver & 40 & 7.05e+03s & 51\%  \\
        Problem in \(\Omega_1\): system assembly & 40 & 41.2s & 0.3\% \\\hline
        Total wallclock time & & 1.39e+04s & 100\%
    \end{tabular}
    \caption{Summary of computational costs of a run of the example \texttt{multidomain\_stokes} on \(1024\) cores.}
    \label{tab:multidomain_stokes}
\end{table}

\subsection{User interface flexibility}
Finally, to demonstrate the ease of implementing new solvers for complex multiphysics problems exploiting the capabilities of \lifex{}, we present the development of a spinodal decomposition model of a binary fluid undergoing shear flow using the advective Cahn-Hilliard equation, a stiff, non-linear, parabolic equation characterized by the presence of fourth-order spatial derivatives \cite{Liu2013}. Spinodal decomposition consists of the separation of a mixture of two or more components to the bulk regions of both, which occurs, for example, when a high-temperature mixture of two or more alloys is rapidly cooled.

The equation was discretized by using FEs in mixed form, thus splitting it into a system of two parabolic-elliptic equations:
\begin{equation*}
\left\{
\begin{aligned}
\frac{\partial c}{\partial t} - \Delta \mu &= 0, & \quad & \text{in } \Omega \times (0, T] = (0, 1)^3 \times (0, T], \\
\mu - \frac{\mathrm{d}f}{\mathrm{d}c}(c) + \lambda \Delta c &= 0, & \quad & \text{in } \Omega \times (0, T], \\
\nabla c\cdot\mathbf{\nu} & = 0, & \quad & \text{on } \partial\Omega \times (0, T], \\
\nabla \mu\cdot\mathbf{\nu} & = 0, & \quad & \text{on } \partial\Omega \times (0, T], \\
c &= 0.63 + 0.01 \sin\left(2000\pi xyz\right), & \quad & \text{in } \Omega \times \{0\},
\end{aligned}
\right.
\end{equation*}
where \(f(c) = 100 c^2(1 - c)^2\).

\begin{figure}
    \centering
    \includegraphics[width=0.5\linewidth]{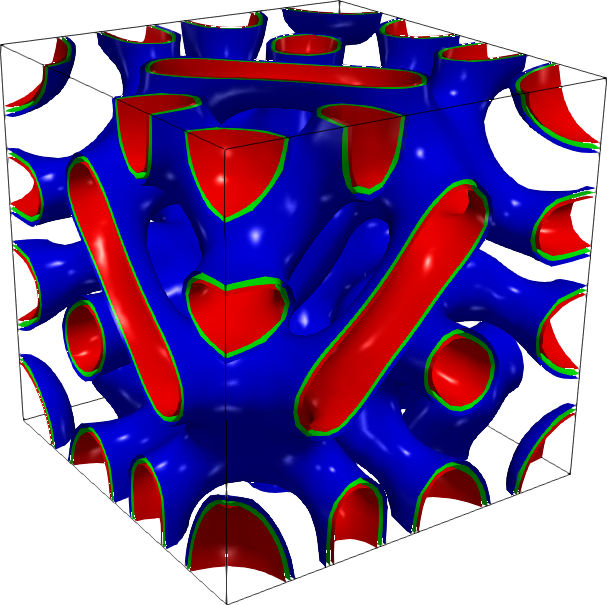}
    \caption{Solution of the Cahn-Hilliard equation implemented in \texttt{Tutorial07}. Isosurfaces corresponding to values of the solution \(c\) equal to \(0.35\) (red), \(0.5\) (green), and \(0.65\) (blue) are shown.}
    \label{fig:cahn_hilliard}
\end{figure}

The problem is discretized in time using a fully implicit scheme through the \texttt{BDFHandler} and linearized using the \texttt{NonLinearSolverHandler} class; at each time step, the Jacobian matrix is computed exploiting automatic differentiation and the associated linear system is solved using GMRES with an AMG preconditioner. Fig.~\ref{fig:cahn_hilliard} shows the steady state solution over the computational domain.

Despite the complexity of the FE formulation of the above problem and the sophisticated numerical schemes adopted for its discretization, the effort involved in implementing such a new solver from scratch in \lifex{} requires writing approximately \(350\) lines of \texttt{C++} code (excluding comments and blank lines). This is a strikingly small number, especially given the vast flexibility the user has in selecting many modeling and numerical features, such as choosing the FE degree, setting up the computational domain and physical parameters, changing the time discretization parameters (including the BDF order), selecting the type of linear solver and preconditioner and related parameters, and so on.

\section{Impact}
The impact and wide applicability of \lifex{} are demonstrated by the high number of journal articles, preprints, conference abstracts, and PhD theses that have already cited it. Computational studies carried out with \lifex{} have appeared in a variety of fields, mostly originating from but not limited to cardiovascular modeling: cardiac electrophysiology \cite{Vergara2022,Piersanti2021,Pagani2021,Stella2020}, cardiac mechanics, electromechanics, and blood circulation \cite{Piersanti2022,Regazzoni2022,Salvador2022,Salvador2021,Dede2021,Quarteroni2022}, fluid dynamics \cite{Zingaro2022b,Zingaro2022a,Fumagalli2022,Fumagalli2020}, FSI \cite{Bucelli2021}, poromechanics coupled with blood perfusion \cite{Barnafi2022,DiGregorio2021}, and hemodynamics in patients affected by COVID-19 \cite{Dede2021a}.

\begin{figure}
    \centering
    \includegraphics[width=0.9\linewidth]{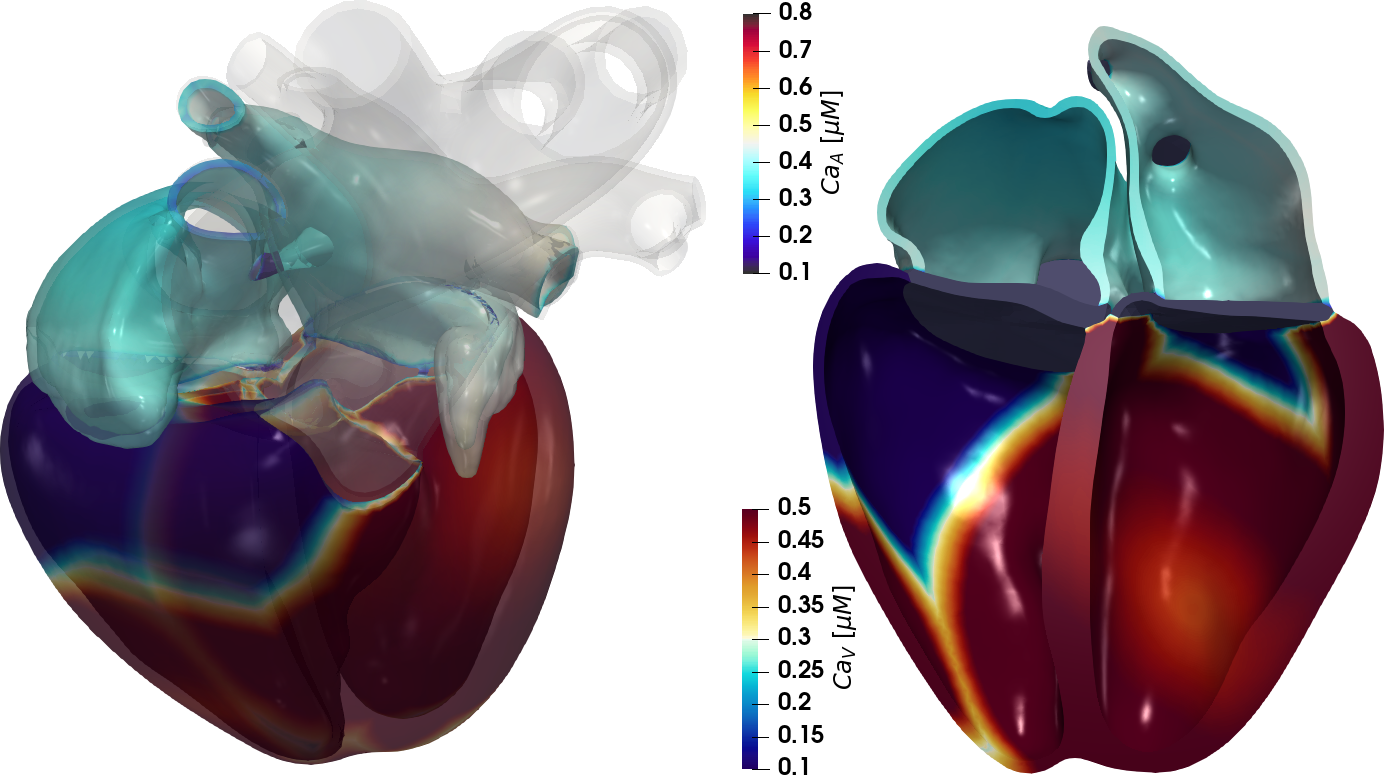}
    \caption{Snapshot of a cardiac electromechanics simulation on a whole-heart geometry during ventricular systole. The color map shows the intracellular concentration of calcium ions in the myocardium. The computational mesh was generated from the Zygote Solid 3D Heart Model \cite{Zygote2014}.}
    \label{fig:heart}
\end{figure}

Other notable computational models relying on \lifex{} involve a comprehensive and biophysically detailed electromechanics of the entire human heart \cite{Fedele2022} (see Fig.~\ref{fig:heart}) and an integrated description of electrophysiology, mechanics, and fluid dynamics in the human left heart \cite{Bucelli2022a,Bucelli2022b}. Other studies oriented toward numerical methods have addressed the development of a high-order, matrix-free solver for cardiac electrophysiology \cite{Africa2022a} and reduced order methods for real-time simulations \cite{cicci2023,Regazzoni2022a,Regazzoni2021}.

\lifex{} provides a fast and stable environment with a gentle learning curve and enables obtaining unmatched results in terms of model reliability, numerical accuracy, and computational efficiency. A comprehensive and up-to-date list of publications making use of \lifex{} can be found at \url{https://lifex.gitlab.io/lifex/publications.html}.

\section{Conclusions}
\lifex{} is a parallel \texttt{C++} library for simulations of multiphysics, multiscale, and multidomain problems based on the \texttt{deal.II} FE core. \lifex{} offers several distinctive features, such as abstract helpers for advanced numerical schemes, a convenient framework to deal with several kinds of coupled problems, and a friendly interface that allows for the implementation of new complex FE solvers or sophisticated numerical schemes with all parameters selectable at run time with an economical number of lines of code. 

The abstraction layer and the software functionalities presented in Sec.~\ref{sec:software_description} show either high parallel speedup or negligible computational overheads. Overall, \lifex{} shows a very high computational efficiency and seamless parallel performance; it is thus an invaluable tool that can be run on diverse architectures, ranging from laptop computers to HPC facilities and cloud platforms.

On the one hand, \lifex{} provides a robust and friendly interface enabling easily accessible and reproducible \textit{in silico} experiments,
without any compromise in terms of computational efficiency and numerical accuracy. On the other, because it is conceived as a research library, \lifex{} can be
exploited by scientific computing experts to address new modeling and numerical challenges in an easily approachable development framework.

Future directions for \lifex{} will include expanding its developer and user bases, maintaining an active and friendly community that welcomes new contributions, and making new advanced features openly available to the wider public (see, for example, \cite{Africa2022,Africa2022b}). In support of these goals, there are a number of technical areas that will open the way to many upcoming developments. First, when moving towards larger numbers of cores or DoFs, new bottlenecks are typically encountered, which will be addressed by carefully profiling all parts of the code base and exploiting more efficient algorithms, such as those based on matrix-free methods \cite{Africa2022a}. Similarly, large computer clusters will likely benefit from GPUs and similar devices in the near future, so targeting new architectures and porting algorithms to different programming models will be of utmost importance \cite{Arndt2021a}. Other possible areas of improvement will target more advanced numerical techniques, such as higher-order or adaptive time advancing schemes, support of multidomain problems discretized over non-conforming meshes (exploiting, for instance, the INTERNODES technique or RBF interpolators \cite{Deparis2014,Deparis2016,Salvador2020}), and \(hp\)-adaptive discretizations to further increase numerical accuracy with a lower computational footprint \cite{Arndt2021a}.

We expect \lifex{} to attract a sizable community of users and developers. Any contribution is highly appreciated, from code commits to bug reports and suggestions for improvement.

As we approach the exascale era that will be dominated by high-end supercomputers, numerical simulations are expected to be one of the main computational workloads \cite{Alowayyed2017}; against this background, \lifex{} offers transparency, accessibility, reproducibility, and reusability of \textit{in silico} experiments, within a flexible, high performance software tool.

\section{Conflict of Interest}
We wish to confirm that there are no known conflicts of interest associated with this publication and that there has been no significant financial support for this work that could have influenced its outcome.

\section*{Acknowledgements}
\label{sec:acknowledgements}
This project has received funding from the European Research Council (ERC) under the European Union's Horizon 2020 research and innovation program (grant agreement No 740132, \href{https://iheart.polimi.it/}{iHEART} - An Integrated Heart Model for the simulation of the cardiac function; P.I. Prof. A. Quarteroni). We acknowledge the CINECA award \texttt{MathBeat} under the ISCRA initiative, for the availability of high performance computing resources and support. The \lifex{} logo was designed by S. Pozzi.

\lifex{} would not have been possible without a large and loyal team working on software development and review, contributing code and fixes or reporting bugs and suggestions: N. Barnafi, L. Bennati, M. Bucelli, L. Cicci, S. Di Gregorio, M. Fedele, I. Fumagalli, S. Pagani, R. Piersanti, F. Regazzoni, M. Salvador, S. Stella, E. Zappon, and A. Zingaro, among many others.

Special thanks go to Profs. A. Quarteroni, L. Dede', L. Formaggia, P. Gervasio, A. Manzoni, C. Vergara, and P. Zunino for many stimulating and inspiring discussions and to L. Paglieri for endless patience and support.

\lifex{} shares the enthusiasm, passion, experience, and dedication to scientific computing brought by several people who contributed to the \texttt{LifeV} library \cite{Bertagna2017}. The name itself was inspired by \texttt{LiFE} (Library of Finite Elements), conceived by Prof. Fausto Saleri.

\bibliographystyle{elsarticle-num}
\bibliography{bibliography.bib}

\end{document}